\documentstyle[aps,epsf,rotating,multicol,amsmath]{revtex}

\preprint{Paper presented at the SLAFES-XV conference}
\begin{document}
\title{Phase Transitions in Confined Antiferromagnets}
\author{A.\ D\'{\i}az-Ortiz and J.\ M.\ Sanchez}
\address{Texas Materials Institute, The University of Texas at
Austin, Austin, TX 78712}
\author{J.\ L.\ Mor\'an-L\'opez}
\address{Instituto de F\'{\i}sica, Universidad Aut\'onoma de San Luis
Potos\'{\i}, 78000 San Luis Potos\'{\i}, S.L.P.\ Mexico}
\date{\today}
\maketitle
\begin{abstract}
Confinement effects on the phase transitions in antiferromagnets are
studied as a function of the surface coupling $v$ and the surface field
$h$ for bcc(110) films. Unusual topologies for the phase diagram are attained
for particular combinations of $v$ and $h$. It is shown that some of the
characteristics of the finite-temperature behavior of the system are
driven by its low-temperature properties and consequently can be explained
in terms of a ground-state analysis. Cluster variation free energies are
used for the investigation of the finite temperature behavior.
\end{abstract}

\pacs{PACS numbers:\ 68.35.Rh; 64.60.Cn}

\begin{multicols}{2}
In recent years the theoretical interest on two-sublattice, uniaxial
antiferromagnets has been renewed\cite{wang1994,trallori1994,%
micheletti1997,trallori1998,micheletti1999}. This interest stems from the
experimental work on Fe/Cr(211) multilayers by Fullerton and
coworkers\cite{fecr}, where an  antiferromagnetic coupling between the Fe
layers is possible for a suitable choice of the Cr layer (11 \AA). For an
even number of layers, the spin-flop phase nucleates at the surface and, as
the external field increases, this surface phase evolves into the bulk
spin-flop phase\cite{wang1994}. The small anisotropy-to-exchange
interaction ratio that characterizes the Fe/Cr(211) system, makes this
particular sort of multilayers amenable to the theoretical modeling. Since
this type of magnetic multilayers are isomorphic to the MnF$_2$ class
antiferromagnets with (100) surfaces, a classical one-dimensional $XY$
model is adequate to model their magnetic properties at low
temperatures\cite{wang1994}. The early work of Mills\cite{mills1968} and
Keffer\cite{keffer1973} on the one-dimensional $XY$ model, that lead to the
identification of the surface spin-flop transition, have been complemented
and extended recently. The surface and finite-size effects on the
ground-state properties of an $XY$ chain have been investigated in terms of
discommensuration transitions\cite{micheletti1997,micheletti1999} and the
analogy between the one-dimensional $XY$ model and Frenkel-Kontorova-type
chains has been elucidated\cite{trallori1998}. These investigations have
set our understanding of the rich magnetic behavior in Fe/Cr multilayers,
where finite-size and surface effects are equally important, on solid
physical grounds. It is appropriate to note that the passage from an
inherently three-dimensional structure, such as the Fe/Cr multilayers, to a
one-dimensional structure is based on the assumption that lateral
fluctuations within each layer can be disregarded with respect to the
interlayer fluctuations. The effect of confinement (surface plus finite
size) in antiferromagnets for which the intralayer fluctuations are
important has also been the subject of previous
investigations\cite{ado-cms97,ado-ssc98,ado-prl98,drewitz,leidl}. 

In this paper we provide a brief survey of a recent study of ground-state
properties of bcc films with [110] surface orientation\cite{ado-prb99} and
relate the previously observed topological features of the phase
diagram\cite{ado-cms97,ado-prl98} to the zero temperature properties of the
system. With these objectives in mind, we consider body-centered
antiferromagnetic Ising films with surfaces oriented in the [110]
direction. In the (110) planes of a bcc structure, each site in one
sublattice has nearest-neighbors in the other sublattice\cite{note1}. For
nearest-neighbor pair interactions, the Hamiltonian is the following:
\begin{multline}
\label{ham}
{\mathcal H}=
J_b\sum_{ij\not\in\text{surf}}\sigma_i\sigma_j +
J_s\sum_{ij\in\text{surf}}\sigma_i\sigma_j\\
-H\sum_{i\in\text{bulk}}\sigma_i-
(h+H)\sum_{i\in\text{surf}}\sigma_i\,,
\end{multline}
where the spin variable $\sigma_i$ takes the value of $+1$ or $-1$
depending if the spin at site $i$ is pointing up or down, respectively. We
have assumed that surface sites, in layers 1 and $N$ for an $N$-layer film,
experience a surface magnetic field $h$ in addition to the external field
$H$. We can think $h$ as the surface perturbation on a highly anisotropic
antiferromagnet (Ising-like) slab, caused by the presence of ferromagnetic
layers in a FM/AFM superlattice. We can also consider $h$ as the
wall-particle interaction in a fluid confined between two parallel plates,
when the usual transformation $p_i=\frac{1}{2} (1 + \sigma_i)$ is used to
cast Hamiltonian (\ref{ham}) into a lattice-gas model. The wall-particle
interaction ($h$) is responsible for the condensation of the liquid phase
at lower chemical potential than it is necessary in the bulk (capillary
condensation)\cite{fn1981,nf1983,evans1990,binder1992,note2}.

Phase equilibrium in confined systems is very sensitive to the boundary
(interface) conditions defined by the surface field $h$ and by the surface
coupling $J_s$\cite{rev-t,rev-e}. In this paper we specialize ourselves to the
case of nearest-neighbor interactions and localized symmetric surface
fields; that is, the field at each surface is the same and acts only at the
surface sites [see Eq.\ (\ref{ham})]. In the remaining of the paper, the
effective pair interactions, the surface field ($h$), and the bulk external
field ($H$) shall be expressed in terms of the bulk AF coupling
($J_b>$0). The ratio of surface to bulk coupling is restricted to positive
values and it is denoted by $v$.

Even when $h$ is zero (neutral boundary conditions), the disruption of the
translation symmetry due to the surfaces results in a ``missing neighbors''
field $h_m$. The surface field $h_m$ is responsible of the inhomogeneities
in the magnetization profile near the surfaces. When Eq.\ (\ref{ham}) is
reinterpreted as a binary-alloy Hamiltonian, the missing-neighbors field,
along with $h$, accounts for the surface segregation phenomenon, i.e., the
enrichment of the surfaces with one component\cite{note2,rev-e,%
surfseg}. In the following we shall consider only the case of $h>0$, since
the results for $h<0$ can be obtained straightforwardly from the symmetry
properties of Hamiltonian (\ref{ham}), as it is discussed next. For zero
surface field, Hamiltonian (\ref{ham}) is invariant under the
transformations $\sigma_i \to -\sigma_i$, $H\to-H$. For neutral boundary
conditions ($h=0$) the ground-state and the finite-temperature properties
of the Hamiltonian are symmetric about $H=0$. A positive value of $h$
breaks this symmetry by favoring the spin-up states at surfaces. Both zero-
and finite-temperature properties of Hamiltonian in Eq.\ (\ref{ham}) become
asymmetric with the applied field $H$. However, for nonzero $h$ the
Hamiltonian is still invariant if we extend the above transformation to
include $h\to-h$.

The selective nature of the surface field has varied consequences on the
properties of Hamiltonian (\ref{ham}), since the equilibrium states are
defined by the competition between the Zeeman and the ordering energies. 
The ordering contribution to the Hamiltonian, first term in the rhs of Eq.\ 
(\ref{ham}), favors AFM structures whereas the Zeeman energy in third sum
of the rhs of (\ref{ham}) promotes FM structures. An additional Zeeman
contribution arises from the surface field [last term in the rhs of
Hamiltonian (\ref{ham})], which competes with the surface ordering
tendencies [second sum in the rhs of (\ref{ham})] to define the equilibrium 
state in the film. Thus, for positive and large values of $H$, applying a
surface field is of little consequence since the stable state is one with
spin-up at the surfaces. For low and negative values of $H$, where
spin-down states are likely to occur, the surface field actually may give
rise to an antiferromagnetic surface state.

\begin{figure}
\epsfysize=3.1in\null
\vskip-3\baselineskip
\centerline{\epsfbox{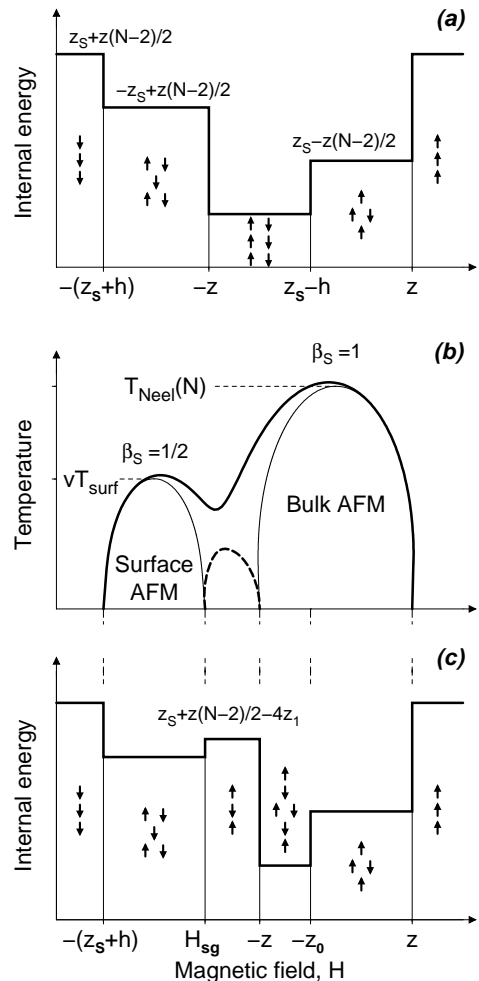}}
\vskip13\baselineskip
\caption{Schematic representation of the ground-state and
finite-temperature phase diagrams for films under intense surface fields.
As a function of the external field $H$, a film of $N$ layers transits
between the ground states (GS) displayed in (a) if $h<h_g$ or in those
showed in (c) if $h>h_g$. For the former case, the corresponding $H$-$T$
critical line is shown in (b) in a thick solid line. (c) For $h>h_g$ a
ferromagnetic (disordered) gap intervenes between $\uparrow \downarrow
/\downarrow / \uparrow \downarrow$ ground state and the zero-magnetization
structure. The characteristic field between the disordered gap and the GS
with surface AFM order is $H_{sg}=z_s-2z_1-h$. For $h$ slightly above
$h_g$, thermal excitations turns the ferromagnetic gap into a disordered
region in the $H$-$T$ plane [dashed line in (b)]. For very intense surface
fields the otherwise connected AFM region splits into two separate critical
curves [thin solid lines in (b)].  See the text for further explanations.}
\label{pd}
\end{figure}

An analysis of the ground states of Hamiltonian (\ref{ham}) singles out the
ground-state (GS) sequence in Fig.\ \ref{pd}(a) as the stable sequence for
large $h$\cite{ado-prb99}. The nomenclature in Fig.\ \ref{pd} is as
follows:\ the intra- and interlayer coordination numbers are represented by
$z_0$ and $z_1$, respectively, with the bulk coordination number expressed
as $z=z_0 +2 z_1$. The parameter $z_s=z_0 v +z_1$ can be regarded as the
surface coordination number but actually accounts for the surface energy
[recall that all quantities in Eq.\ (\ref{ham}) are normalized to
$J_b$]. Label $\uparrow\downarrow/ \downarrow / \uparrow\downarrow$ stands
for a $N$-layer film with AFM surfaces and down magnetization in the
remaining $(N-2)$ layers. 

From Fig.\ \ref{pd}(a) one can see that a GS structure with AFM surface
coexists with a ferromagnetic bulk for $H\in(-z_s-h, -z)$, while the
contrary occurs for $H\in(z_s-h,z)$. In between, i.e.\ for $H\in(-z,z_s-h)$,
the GS is AFM in both the surfaces and the bulk. The film \hphantom{displays}
%
\noindent displays an ordered, compact domain from $H=-(z_s+h)$ to
$H=z$. In Fig.\ \ref{pd}(b), in thick solid line, we show the critical
curve (schematic) in the $H$-$T$ plane associated with the GS sequence of
Fig.\ \ref{pd}(a).

For negative values of the external field ($H\lesssim -z$), the surface
field favors the AFM ordering at the surfaces but also promotes the
decoupling of the surface layers from the rest. Therefore, the inner layers
closely behave as a $(N-2)$-layer film with neutral boundary conditions. In
Fig.\ \ref{pd}(b) with thin lines and appropriately shifted, we have
plotted the corresponding critical curves for a 2D square lattice (left)
and the corresponding $(N-2)$-layer film at $h=0$ (right). Observe that the
shoulder shows a maximum temperature $\sim vT_{\text{surf}}$, where
$T_{\text{surf}}$ is the N\'eel temperature of the 2D square lattice.

The ground-state sequence in Fig.\ \ref{pd}(a) becomes unstable upon an
increment in the surface field, and the GS sequence of Fig.\ \ref{pd}(c) is
then adopted by the film. Note the a disordered gap (GS $\uparrow/
\downarrow / \uparrow$) and a new zero-magnetization ground structure
($\uparrow / \downarrow / \uparrow\downarrow / \downarrow / \uparrow$:\ an
up-magnetization state at the surfaces, subsurface layers down
magnetization and the rest in the AFM state) appear in lieu of the homogeneous
AFM-GS structure. It can be shown\cite{ado-prb99} that the transition
between GS sequences, from the sequence in Fig.\ \ref{pd}(a) to the one in
Fig.\ \ref{pd}(c), occurs at a surface field value $h_g$ given by
\begin{equation}
\label{hg}
h_g=z_0v+(z_0+z_1).
\end{equation}
For a surface field slightly above $h_g$, the disordered gap transforms
itself, via thermal excitations, into a disordered region in the plane
$H$-$T$, right in the middle of the ordered region [Fig.\ \ref{pd}(b),
dashed line]. In other words, the phase diagram is composed by two critical
lines [dashed and thick solid lines in Fig.\ \ref{pd}(b)]. Upon
high-temperature cooling and for $H\in(z_s-2z_1-h,-z)$, the system
undergoes a phase transition from the high-temperature disordered state to
an AFM state. A further decrease in temperature drives the system into a
low-temperature disordered state.

Intense surface fields increase the disordered gap at zero temperature and,
as a consequence, the height of the associated disordered region rises. At
$h=h_s$, the AFM domain splits into the surface and the bulk critical
curves. In Fig.\ \ref{pd}(b) with thin lines, the critical curves
associated with the surfaces and the bulk are presented for $h>h_s$. It is
shown in Ref.\ \onlinecite{ado-prb99} that the splitting point corresponds
to a saddle point in the Hessian of the free energy as a function of $T$
and $H$\cite{note3}. We have used the pair approximation of the
cluster-variation method (CVM)\cite{cvm} to evaluated the
finite-temperature properties of Hamiltonian (\ref{ham}). Previous work
have shown that for nonfrustrated lattices, such as the bcc and simple
cubic, the PA gives reliable results for the qualitative aspects of the
phase equilibrium in restricted geometries\cite{ado-cms97,ado-ssc98,%
ado-prl98}.  

In a sense, the splitting value of the surface field, $h_s$, represents at
finite temperatures the role of $h_g$. Both characteristic values of the
surface field $h_g$ and $h_s$, are the answer for the following question:\
How intense need the surface field be, in order to split the otherwise
compact AFM domain, into separate surface and bulk ordered regions? At zero
Kelvin, the answer is independent of the film thickness:\ When surface
field reaches the value of $h_g=z_0v +(z_0+z_1)$, the surface splits from
the bulk independently of the number of layers. At finite temperatures, the
answer is more involved since now thermal excitations enhance the coupling
between the bulk and
\begin{figure}
\epsfxsize=2.9in\null
\centerline{\epsfbox{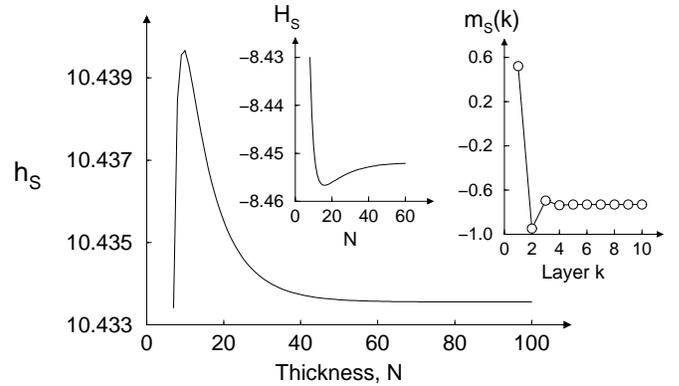}}
\caption{Surface and bulk fields corresponding to the splitting point $h_s$ 
and $H_s$ (left inset), respectively, as a function of the thickness of the
film. A magnetization profile for $N=100$ film at $h=h_s$ is also shown in
the right inset. Calculations were performed in the pair approximation of
the CVM for bcc(110) films with $v=1$.}
\label{sp}
\end{figure}
\noindent the surface layers. Figure \ref{sp} shows two regimes
of behavior for $h_s$ as a function of the film thickness:\ for thin films
the value of $h_s$ increases with $N$ while the contrary occurs for thick
films. This peculiar behavior of $h_s(N)$ results from the balance between
the surface Zeeman energy and the ordering energy. For very thin films the
surface Zeeman energy easily overcomes the contribution of a bulk made of a
few layers. In this regimen, increasing the thickness in the film is
equivalent to enhancing the bulk contribution to the free energy. Thus, it
is necessary to apply more intense surface fields to split the surfaces
from the bulk.

For very thick films the splitting value of the surface field shows
virtually no change as the thickness in the film is reduced. Near the
splitting point, the magnetization profile decays very fast towards the
bulk state as we move from the surfaces to the inner layers (see inset in
Fig.\ \ref{sp}). The surfaces are too far away to affect each other and,
therefore, $h_s$ corresponds to the semiinfinite value of the surface field
$h_s^\infty$. However, if the film thickness is reduced enough ($N\sim50$
in Fig.\ \ref{sp}), the perturbation introduced by the surfaces reaches the 
middle layers. The interplay between the surface and the finite-size
effects is reflected as an increment in the value of $h_s$ as the thickness
is decreased.

In summary, we have shown that the rich magnetic behavior, previously
reported in AFM thin films\cite{ado-cms97,ado-prl98}, is directly related
to the ground state properties of the films. We focused on the
thermodynamic behavior for intense surface fields, since in that
case the otherwise compact antiferromagnetic regions splits into
surface- and bulk-driven critical curves. In the bulk-driven critical curve,
the surfaces are less ordered than the layers in the bulk. On the other hand,
along the line of phase transitions driven by the surface, the bulk layers
are less ordered than the surfaces. For surface fields such as $h>h_s>h_g$,
in which the critical curve is well separated into the bulk- and
surface-driven AFM regions, the surface order parameter vanishes with
exponent $\beta_s$, which in the mean field approximation used here, equals
1 for the bulk-driven critical line, whereas $\beta_s=\frac{1}{2}$ for the
surface-driven phase transitions.  However, our preliminary results show
that {\em even for $h<h_g$ the surface exponent changes from $\beta_s=1$ to
$\beta_s=\frac{1}{2}$ as the external field is varied from positive to
negative values}. The investigation of the critical behavior will be
considered in the future.

\acknowledgments
This work was sponsored by Consejo Nacional de Ciencia y Tecnolog\'{\i}a
(CONACyT), Mexico, through grant G-25851-E. AD-O gratefully acknowledges the
financial support from CONACyT through the Post Doctoral Fellowships
Program.

\end{multicols}

\begin{thebibliography}{99}
\bibitem{note1} In Refs.\ \onlinecite{drewitz} and \onlinecite{leidl} a
(110) surface in a bcc antiferromagnet is called symmetry-preserving while
(100), where each layer belongs to one of the two sublattices, is
denoted by symmetry-breaking orientation.
\bibitem{note2} In the context of binary alloys, the surface field $h$ can
also be regarded as proportional to the difference in the enthalpies of
formation of the pure elements.
\bibitem{note3} The matrix of second derivatives of the free energy with
respect of the long-range order parameters is called the Hessian of the
free energy. The Hessian is a function of the external fields and the order
parameters.
\end{thebibliography}
\end{document}